\documentclass[11pt]{article}
\usepackage{axodraw}
\usepackage{graphicx}
\usepackage{enumerate}
\newcommand{\be}{\begin{equation}}
\newcommand{\ee}{\end{equation}}
\newcommand{\ba}{\begin{eqnarray}}
\newcommand{\ea}{\end{eqnarray}}

\def\lsim{\raise0.3ex\hbox{$\;<$\kern-0.75em\raise-1.1ex\hbox{$\sim\;$}}}
\def\gsim{\raise0.3ex\hbox{$\;>$\kern-0.75em\raise-1.1ex\hbox{$\sim\;$}}} 

\newcommand{\mx}{\left[\begin{array}}
\newcommand{\finmx}{\end{array}\right]} 
\newcommand{\mxp}{\left(\begin{array}} 
\newcommand{\finmxp}{\end{array}\right)} 
\def\beq{\begin{equation}}
\def\eeq{\end{equation}}
\def\bea{\begin{eqnarray}}
\def\eea{\end{eqnarray}}

\def\mathbf#1{\hbox{\bf #1}}
\def\textrm#1{\hbox{#1}}

\def\lsim{\raise0.3ex\hbox{$\;<$\kern-0.75em\raise-1.1ex\hbox{$\sim\;$}}}
\def\gsim{\raise0.3ex\hbox{$\;>$\kern-0.75em\raise-1.1ex\hbox{$\sim\;$}}}
\newcommand {\ignore}[1]{}

\begin{document}
\vspace*{-1in}
\renewcommand{\thefootnote}{\fnsymbol{footnote}}
\begin{flushright}
\texttt{
} 
\end{flushright}
\vskip 5pt
\begin{center}
{\Large{\bf LFV Constraints on the Majorana Mass Scale in mSUGRA}\footnote{Talk presented by A.~Redelbach at SUSY02, 10th International Conference on \it{Supersymmetry and Unification of Fundamental Interactions}\rm, 17-23/06/02, DESY, Hamburg}}
\vskip 25pt

{\sf 
F. Deppisch, 
H. P\"as, 
A. Redelbach, 
R. R\"uckl
}
\vskip 10pt
{\it \small Institut f\"ur Theoretische Physik und Astrophysik\\
Universit\"at W\"urzburg\\ D-97074 W\"urzburg, Germany}\\

\vskip 20pt

{\sf 
Y. Shimizu
}
\vskip 10pt
{\it \small Department of Physics \\ Nagoya University\\ Nagoya, 464-8602, Japan}\\

\vskip 20pt

{\bf Abstract}
\end{center}

\begin{quotation}
{\small 
We discuss constraints on the right-handed Majorana mass scale
$M_R$ of the SUSY see-saw model in the mSUGRA framework. 
The sensitivity of radiative lepton-flavor violating decays on \(M_R\) is compared with the reach in lepton-flavor violating channels at a future linear collider. 
}
\end{quotation}

\vskip 20pt  

\setcounter{footnote}{0}
\renewcommand{\thefootnote}{\arabic{footnote}}


\section{Introduction}

While lepton-flavor violating processes are suppressed due to the small 
neutrino masses if only right-handed neutrinos are added to the Standard Model
\cite{petcov}, in supersymmetric models new sources of lepton-flavor violation (LFV) exist \cite{othertalks}. In this report, the focus is on virtual effects of the massive neutrinos in the renormalization group running of the slepton mass and 
trilinear coupling matrices that give rise to LFV.
Adopting the SUSY see-saw model we investigate ways to probe and extract the Majorana mass scale \(M_R\) from measurements of the branching ratios 
$Br(l_{i}\rightarrow l_{j}\gamma)$, $l_{i}=e,\mu,\tau$ in low-energy experiments.
This is compared to the sensitivity of the high-energy processes \(e^+e^-\rightarrow l_i^-l_j^+ + /\!\!\!\!E\) at an electron-positron linear collider.

\section{Induced LFV in the SUSY see-saw framework}

The supersymmetric see-saw mechanism with three right-handed neutrino singlet fields leads to the following mass matrix for the light neutrinos \cite{Casas:2001sr}:
\beq\label{eqn:SeeSawFormula}
M_\nu = m_D^T M^{-1} m_D = Y_\nu^T M^{-1} Y_\nu (v \sin\beta )^2.
\eeq
Here \(Y_\nu\) denotes the matrix of neutrino Yukawa couplings and \(m_D=Y_\nu \langle H_2^0 \rangle\) the corresponding Dirac mass matrix with
\(\langle H_2^0 \rangle = v\sin\beta\) being the appropriate Higgs v.e.v., \(v=174\)~GeV and \(\tan\beta =
\frac{\langle H_2^0\rangle}{\langle H_1^0\rangle}\).
Light neutrino masses are naturally obtained if the mass scale \(M_R\) of the
 Majorana matrix \(M\) of right-handed neutrinos is much larger than the electroweak scale of the Dirac mass matrix \(m_D\).
The matrix $M_{\nu}$ is diagonalized by the unitary MNS matrix \(U\),
\beq\label{eqn:NeutrinoDiag}
U^T M_\nu U = \textrm{diag}(m_1, m_2, m_3),
\eeq
in order to obtain the mass eigenvalues \(m_i\) of the light neutrinos.
The matrix  $U$ and the masses \(m_i\) are constrained by neutrino experiments \cite{Deppisch:2002vz}.

On the other hand, the massive neutrinos give rise to virtual corrections to the slepton mass matrices that are responsible for lepton-flavor violating effects.
These corrections depend on the product $Y_\nu^\dagger Y_\nu$, where \cite{Casas:2001sr}
\begin{eqnarray}\label{eqn:yy}
 Y_\nu = \frac{\sqrt{M_R}}{v\sin\beta}\cdot R\cdot \textrm{diag}(\sqrt{m_1},\sqrt{m_2},\sqrt{m_3})\cdot U^\dagger.
\end{eqnarray}
Here, we work in the basis where the charged lepton Yukawa couplings are diagonal and assume degenerate right-handed Majorana masses.
If the unknown complex orthogonal matrix \(R\) is real, a case which is sufficient to make our points, it drops out from the product $Y_\nu^\dagger Y_\nu$.
For hierarchical (a) and degenerate (b) neutrinos one then obtains
\begin{eqnarray}
\mbox{(a) }
 \left(Y_{\nu}^{\dagger}Y_{\nu}\right)_{ij} &\approx&
\frac{M_{R}}{v^{2}\sin^{2}\beta}
 \left(\sqrt{\Delta m^{2}_{12}}U_{i2}U_{j2}^{*} +
      \sqrt{\Delta m^{2}_{23}}U_{i3}U_{j3}^{*}\right) \label{llcorrectionhier} \\
\mbox{(b) } \left(Y_{\nu}^{\dagger}Y_{\nu}\right)_{ij} &\approx&
\frac{M_{R}}{v^{2}\sin^{2}\beta}
\left(m_{1}\delta_{ij}+\left(
\frac{\Delta m^{2}_{12}}{2m_1}U_{i2}U_{j2}^{*} +
\frac{\Delta m^{2}_{23}}{2m_1}U_{i3}U_{j3}^{*}\right)\right) \label{llcorrectiondeg},
\end{eqnarray}
where \(\Delta m_{ij}^2=m^2_j-m^2_i\) and, in (b), \(m_1^2\gg\Delta m^2_{23}\gg\Delta m^2_{12}\).

Using the results from neutrino experiments as an input at the low scale \(\mathcal{O}(M_Z)\), the neutrino Yukawa couplings are evolved to the unification scale \(M_X\). Then, the slepton mass matrix is run from \(M_X\) to the electroweak scale assuming the mSUGRA universality conditions,
\begin{equation}
m^{2}_{L}=m_{0}^{2}\mathbf{1},\qquad m^{2}_{R}=m_{0}^{2}\mathbf{1},
\qquad A_{l}=A_{0}Y_{l},
\end{equation}
where $m_{0}$ is the common soft SUSY-breaking scalar mass and $A_{0}$ the trilinear coupling. 
This evolution generates off-diagonal terms involving \(Y_\nu(M_X)\) which
in leading-log approximation are given by \cite{Hisano:1999fj}
\begin{eqnarray}\label{eq:rnrges}
  \delta m_{L}^2 &=& -\frac{1}{8 \pi^2}(3m_0^2+A_0^2)(Y_\nu^\dag Y_\nu) 
\ln\left(\frac{M_X}{M_R}\right) \\
  \delta m_{R}^2 &=& 0  \\
  \delta A &=& -\frac{3 A_0}{16\pi^2}(Y_l Y_\nu^\dag Y_\nu) \ln\left(\frac{M_X}{M_R}\right).
\end{eqnarray}

\section{LFV at low energies}

At low energies, the flavor off-diagonal terms in the slepton mass matrix induce radiative decays such as \(l_i\rightarrow l_j \gamma\).
The effective Lagrangian for $l_{i}^{-}\rightarrow l_{j}^{-}\gamma$ is given by
\cite{Carvalho:2001ex}
\begin{equation}
\mathcal{L}_{eff}=\frac{e}{2}\bar{l}_{j}\sigma_{\alpha \beta}F^{\alpha \beta}\left(A_{L}^{ij}P_{L}+A^{ij}_{R}P_{R}\right)l_{i},\label{Leff}
\end{equation}
where $F^{\alpha\beta}$ is the electromagnetic field strength tensor, $\sigma_{\alpha\beta}=\frac{i}{2}\left[\gamma_{\alpha},\gamma_{\beta}\right]$, $i$ and $j$ are flavor indices, and $P_{R,L}=\frac{1}{2}(1\pm \gamma_{5})$ are the helicity projection operators.
The coefficients $A^{ij}_{L,R}$ are calculated from the photon penguin diagrams shown in Fig.~1 with charginos/sneutrinos or neutralinos/charged sleptons in the loop.
\begin{center}
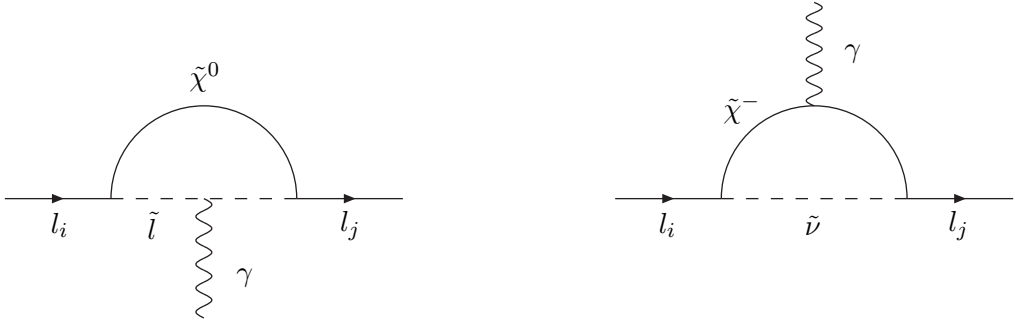
\begin{figure}[h!] 
\begin{picture}(400,135)(-200,-50)
\ArrowLine(-190,0)(-150,0)
\Text(-170,-10)[]{$l_{i}$}
\DashLine(-150,0)(-80,0){5}
\Text(-135,-10)[]{$\tilde{l}$}
\Photon(-115,0)(-115,-45){3}{5}
\Text(-100,-30)[]{$\gamma$}
\ArrowLine(-80,0)(-40,0)
\Text(-60,-10)[]{$l_{j}$}
\CArc(-115,0)(35,0,180)
\Text(-115,46)[]{$\tilde{\chi}^{0}$}
\ArrowLine(40,0)(80,0)
\DashLine(80,0)(150,0){5}
\ArrowLine(150,0)(190,0)
\Text(60,-10)[]{$l_{i}$}
\Text(115,-10)[]{$\tilde{\nu}$}
\Photon(115,35)(115,74){3}{5}
\Text(130,55)[]{$\gamma$}
\Text(170,-10)[]{$l_{j}$}
\CArc(115,0)(35,-360,-180)
\Text(88,34)[]{$\tilde{\chi}^{-}$}
\end{picture} 
\caption{\label{lfv_lowenergydiagrams} Diagrams for $l_{i}^{-}\rightarrow l_{j}^{-}\gamma$ in the MSSM}
\end{figure} 
\end{center}
From (\ref{Leff}) one obtains the decay rate \cite{Carvalho:2001ex}
\begin{equation}
\Gamma\left(l_{i}^{-}\rightarrow l_{j}^{-}\gamma\right)=\frac{\alpha}{4}m_{l_{i}}\left(\left| A_{L}^{c}+A_{L}^{n}\right|^{2}+\left|A_{R}^{c}+A_{R}^{n}\right|^{2}\right),
\end{equation}
where the superscript \(c\,(n)\) refers to the chargino (neutralino) diagram of Fig.~1, while flavor indices \(i,j\) are omitted.
Note that the coefficients \(A_{R,L}^{c,n}\) are roughly proportional to \(m_{l_i}^2\frac{\delta m_{L,R}^2}{m_S^4}\), where \(m_S\) denotes the typical mass scale of supersymmetric particles in the loop.

\section{LFV in high energy processes}

A favourable channel at high energies is represented by the process
$e^+e^- \to \tilde{l}_\alpha^+\tilde{l}^-_\beta\to 
l_i^-l^+_j\tilde{\chi}^0_a\tilde{\chi}^0_b$.
From the Feynman graphs shown in Fig.~2 one can see that lepton-flavor violating couplings appear both in production and decays vertices.
The helicity amplitudes \(M_{\alpha\beta}\) for the pair production of \(\tilde{l}_{\alpha}^+\) and \(\tilde{l}_\beta^-\), as well as the decay amplitudes \(M_\alpha\) and \(M_\beta\) are given explicitly in \cite{inprep}. 
The amplitude squared can be written as
\begin{eqnarray}
  |M|^2\!\!&=&\!\!\sum_{\alpha\beta\gamma\delta}(M_{\alpha\beta} M_{\gamma\delta}^{*})(M_\alpha M_\gamma^{*})(M_\beta M_\delta^{*})C_{\alpha\gamma}
     \frac{\pi}{2\overline{m\Gamma}_{\alpha\gamma}}C_{\beta\delta}\frac{\pi}{2\overline{m\Gamma}_{\beta\delta}} \nonumber\\
          &\quad& \times \left(\delta(p_3^2-m_{\tilde{l}_\alpha}^2)+\delta(p_3^2-m_{\tilde{l}_\gamma}^2)\right) 
                         \left(\delta(p_4^2-m_{\tilde{l}_\beta}^2)+\delta(p_4^2-m_{\tilde{l}_\delta}^2)\right)\label{full_M_squared},
\end{eqnarray}
where the sum runs over all internal sleptons, and
\begin{eqnarray}
  C_{\alpha\gamma}&=&\frac{1}{1+i\frac{\Delta \tilde{m}_{\alpha\gamma}^2}{2\overline{m\Gamma}_{\alpha\gamma}}},\nonumber\\
\overline{m\Gamma}_{\alpha\gamma}&=&\frac{1}{2}(m_{\tilde{l}_\alpha}\Gamma_{\tilde{l}_\alpha}+m_{\tilde{l}_\gamma}\Gamma_{\tilde{l}_\gamma}), \qquad  \Delta \tilde{m}_{\alpha\gamma}^2=m_{\tilde{l}_\alpha}^2-m_{\tilde{l}_\gamma}^2 \label{flavor_correlations}.
\end{eqnarray}
Obviously, the narrow width approximation has been used here in the slepton propagators.
\begin{center}
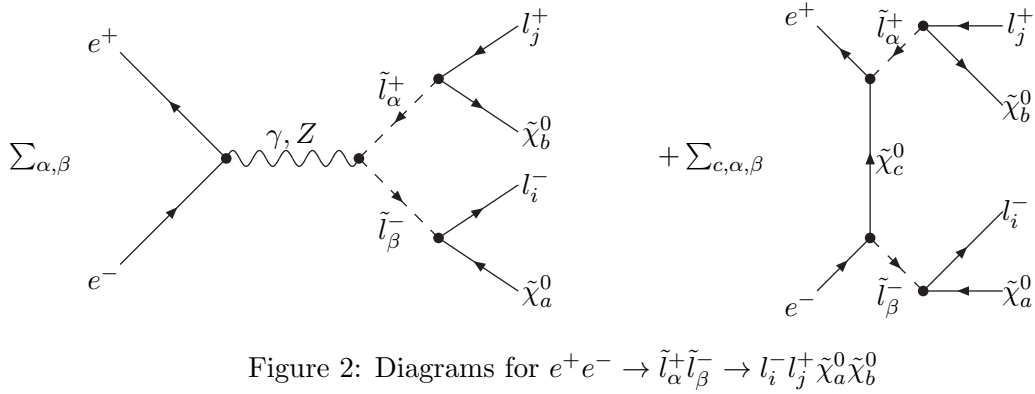
\begin{figure}[h!]
\begin{picture}(185,50)(-50,48)  \Text(-20,50)[c]{\( \sum_{\alpha,\beta} \)}
                           \ArrowLine(50,50)(10,90)        \Text(10,95)[r]{$e^{+}$}
                           \ArrowLine(10,10)(50,50)        \Text(10,5)[r]{$e^{-}$}              \Vertex(50,50){2}
                           \Photon(50,50)(100,50){3}{5}           \Text(75,54)[b]{\(\gamma,Z\)} \Vertex(100,50){2}
                           \DashArrowLine(130,80)(100,50){5}  \Text(113,70)[b]{\(\tilde{l}^{+}_{\alpha}\)}  \Vertex(130,80){2}
                           \ArrowLine(160,100)(130,80)        \Text(162,100)[l]{\(l_{j}^{+}\)}
                           \ArrowLine(130,80)(160,60)        \Text(162,60)[l]{\(\tilde{\chi}^{0}_{b}\)}
                           \DashArrowLine(100,50)(130,20){5}  \Text(113,30)[t]{\(\tilde{l}_{\beta}^{-}\)}  \Vertex(130,20){2} 
                           \ArrowLine(130,20)(160,40)        \Text(162,40)[l]{\(l^{-}_{i}\)}
                           \ArrowLine(160,0)(130,20)        \Text(162,0)[l]{\(\tilde{\chi}_{a}^{0}\)}
                         \end{picture}
                         \begin{picture}(-40,50)(-135,48)\Text(-40,50)[c]{\(+\sum_{c,\alpha,\beta}\)}
                           \ArrowLine(20,80)(0,100)   \Text(0,105)[r]{$e^{+}$}
                           \ArrowLine(0,0)(20,20)     \Text(0,-5)[r]{$e^{-}$} 
                           \ArrowLine(20,20)(20,80)   \Text(22,50)[l]{\(\tilde\chi_c^{0}\)} \Vertex(20,20){2} \Vertex(20,80){2}
                           \DashArrowLine(40,100)(20,80){5} \Text(28,93.5)[b]{\(\tilde{l}_{\alpha}^{+}\)} \Vertex(40,100){2}
                           \ArrowLine(70,100)(40,100)  \Text(72,100)[l]{\(l_{j}^{+}\)}
                           \ArrowLine(40,100)(70,70) \Text(70,70)[l]{\(\tilde{\chi}_{b}^{0}\)}
                           \DashArrowLine(20,20)(40,0){5} \Text(28,5)[t]{\(\tilde{l}^{-}_{\beta}\)}    \Vertex(40,0){2}
                           \ArrowLine(40,0)(70,30)  \Text(70,0)[l]{\(\tilde{\chi}_{a}^{0}\)}
                           \ArrowLine(70,0)(40,0) \Text(70,30)[l]{\(l^{-}_{i}\)}
\end{picture}
\vspace{2cm}
\caption{Diagrams for $e^+e^-\to \tilde{l}_\alpha^+\tilde{l}^-_\beta \to l_i^-l^+_j\tilde{\chi}^0_a\tilde{\chi}^0_b$}\label{e+e-_diags}
\end{figure}
\end{center}
\vspace{-1.1cm}
\section{Numerical Results}

\begin{figure}[t!]
\setlength{\unitlength}{1cm}
\begin{minipage}[t!]{7.5cm}
\includegraphics[clip,scale=0.5]{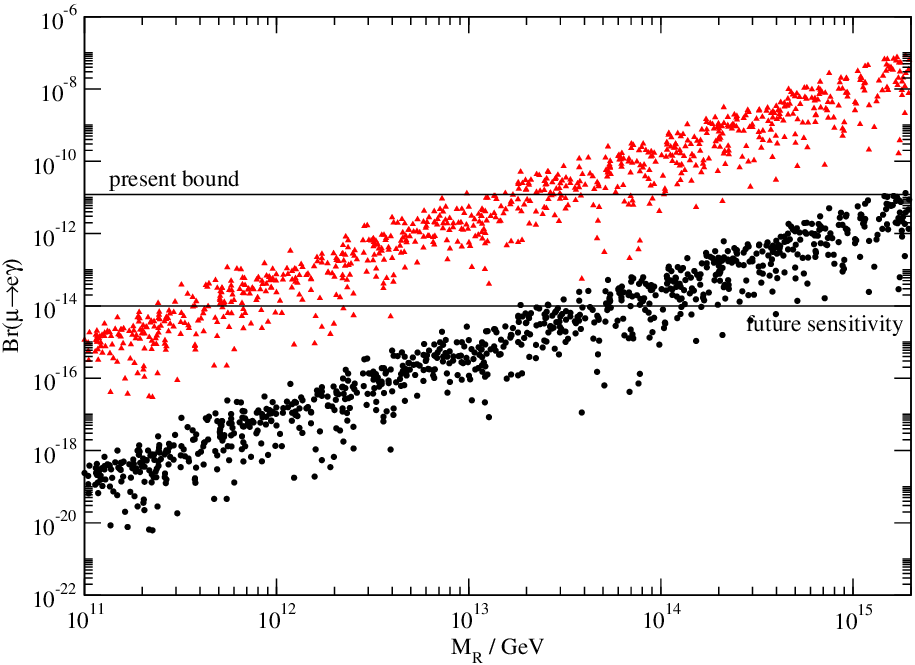}
\caption{Branching ratio of \(\mu \rightarrow e\gamma\) for a hierarchical neutrino spectrum and future neutrino uncertainties in the mSUGRA scenarios giving the largest (L, upper) and the smallest (H, lower) effect.}
     \label{fig:emu_hier_fut}
\end{minipage}\hfill
\begin{minipage}[t!]{7.5cm}
\includegraphics[clip,scale=0.5]{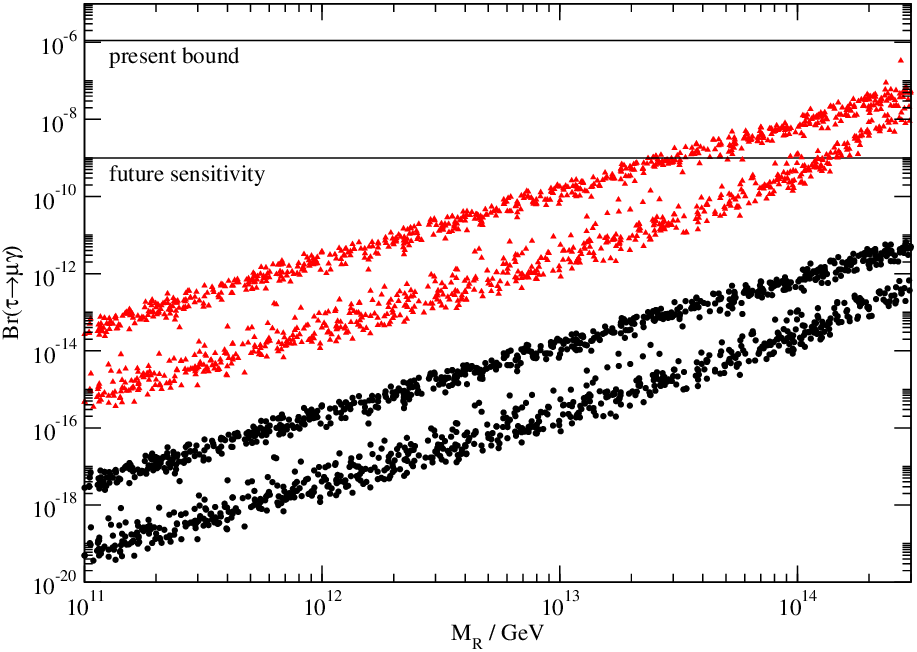}
\caption{Branching ratio of \(\tau \rightarrow \mu\gamma\) in the same mSUGRA scenarios as considered in Fig.~3 for degenerate (lower) and hierarchical (upper) neutrino spectra.}
     \label{fig:mutau_all}
\end{minipage}
\end{figure}
\begin{figure}
\setlength{\unitlength}{1cm}
\begin{minipage}[b!]{7.5cm}
\includegraphics[clip,scale=0.5]{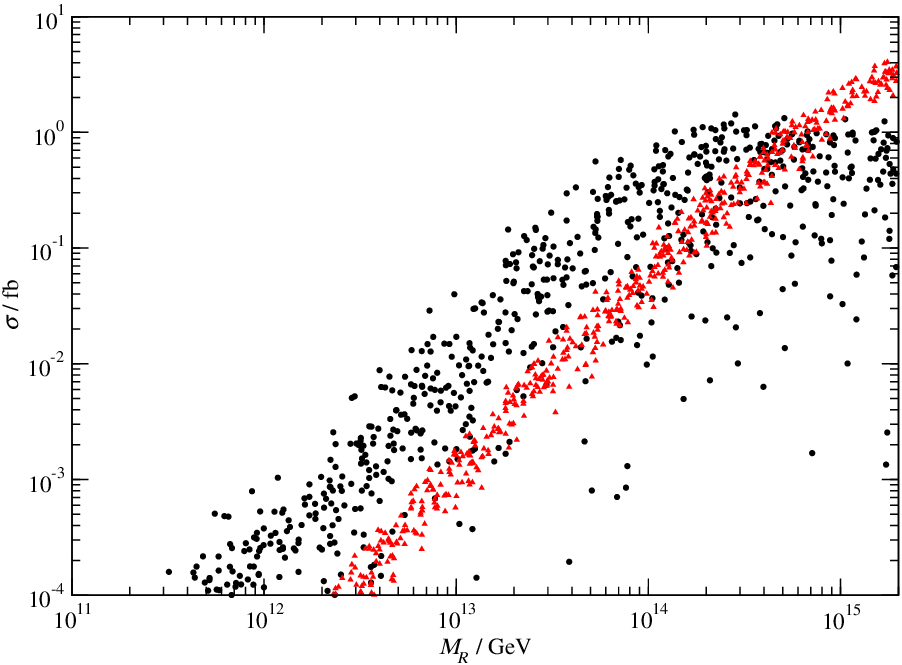}
\caption{Cross-section for \(e^+e^- \to e^-\mu^+ +2\tilde\chi_1^0\) (circles) and \(e^+e^- \to \mu^-\tau^+ +2\tilde\chi_1^0\) (triangles) in scenario B at \(\sqrt{s}=500\) GeV for hierarchical neutrinos.}
\end{minipage}\hfill
\begin{minipage}{7.5cm}
\includegraphics[clip,scale=0.5]{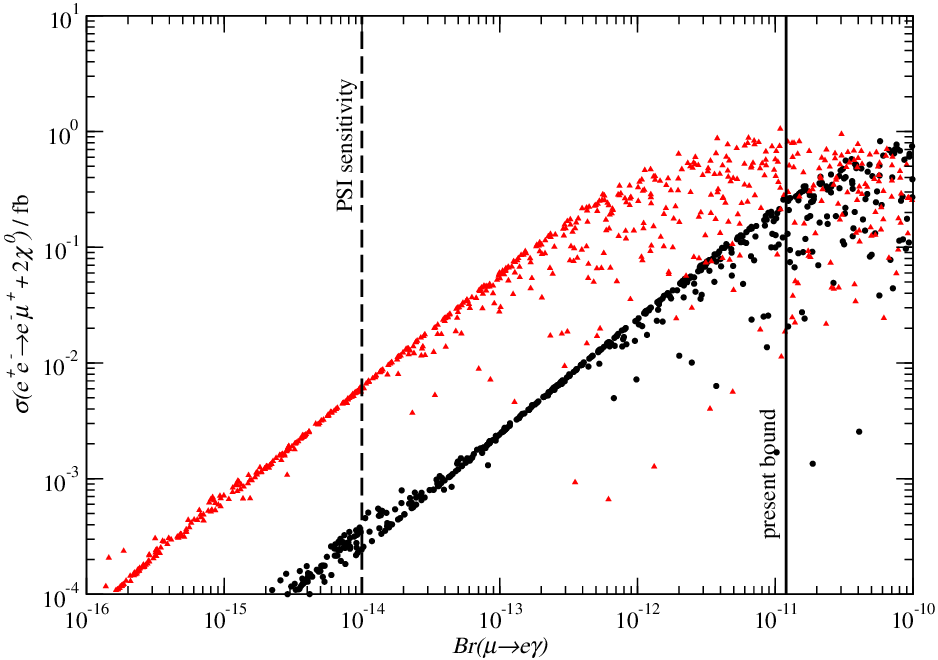}
\caption{Correlation between \(Br(\mu\to e\gamma)\) and   \(\sigma(e^+e^- \to e^-\mu^+ +2\tilde\chi_1^0)\) in scenarios B (circles) and C (triangles) at \(\sqrt{s}=500\) GeV for hierarchical neutrinos.}
\end{minipage}
\end{figure}

For numerical investigations we consider the mSUGRA benchmark scenarios proposed in \cite{Battaglia:2001zp} for linear collider studies.
Only the benchmark scenarios B, C, G, and I have sleptons light enough to be pair-produced in \(e^+e^-\) collisions at c.m. energies \(\sqrt{s}=500\)~GeV and 800~GeV. On the other hand, in the case of radiative lepton decays with SUSY particles in the loop, all thirteen mSUGRA scenarios proposed in \cite{Battaglia:2001zp} are interesting.

For the neutrino input we take the global three neutrino LMA fit given in \cite{Gonzalez-Garcia:2001sq} and vary the parameters within the 90\% CL intervals.
In order to demonstrate possible future improvements, we also show results for the smaller uncertainty intervals expected in future experiments \cite{Deppisch:2002vz}.
Finally, the Majorana mass scale \(M_R\) is treated as a free parameter.

The dependence of the decays \(\mu\rightarrow e\gamma\) and \(\tau\rightarrow \mu\gamma\) on \(M_R\) is displayed in Figs.~3 and 4 for the mSUGRA scenarios L and H, which lead to the largest and smallest branching ratios, respectively. The present experimental bounds still allow \(M_R\) to be as large as \(2\cdot 10^{15}\)~GeV, while the sensitivity of the new PSI experiment on \(\mu\rightarrow e\gamma\) will probe \(M_R\) below \(5 \cdot 10^{14}\)~GeV in all mSUGRA scenarios considered.
By comparing Fig.~3 to Fig.~4, one can see that the branching ratio of $\mu\rightarrow e \gamma$ reflects more strongly the uncertainties in the neutrino parameters than the branching ratios of $\tau\rightarrow \mu \gamma$. This is because \(Br(\mu\rightarrow e \gamma)\) also depends on the small mixing angle \(\theta_{13}\) which is badly known, whereas  \(Br(\tau\rightarrow \mu \gamma)\)  
mainly depends on the large angle \(\theta_{23}\).
More quantitatively, while a measurement
of $Br(\mu\rightarrow e\gamma)$ would only allow to constrain \(M_R\) within a factor of 10-100, in a given mSUGRA scenario, a measurement $Br(\tau \rightarrow \mu \gamma)$ would determine \(M_R\) within a factor of 2.
On the other hand, \(\mu \rightarrow e \gamma\) is sensitive to lower values of \(M_R\) than \(\tau \rightarrow \mu \gamma\). 
  
Fig.~5 shows the cross-section for \(e^+e^-\rightarrow e^-\mu^++2\tilde{\chi}^0_1\) and \(e^+e^-\rightarrow \mu^-\tau^++2\tilde{\chi}^0_1\) versus \(M_R\) at the c.m. energy \(\sqrt{s}=500\)~GeV. Since the channel \(e^-\tau^++2\tilde{\chi}_1^0\) is strongly suppressed by the small mixing angle \(\theta_{13}\), it is more difficult to observe and not shown here.
An important point is the strong correlation between LFV in the radiative decays and in  high-energy slepton pair production. This is illustrated in Fig.~6 for \(Br(\mu\to e \gamma)\) and \(e^+e^-\rightarrow e^-\mu^++2\tilde{\chi}_1^0\).
Because of this correlation the neutrino uncertainties almost drop out in such a comparison, while the dependence on the mSUGRA scenarios remains.

Finally it is interesting to note that in the case of degenerate neutrino 
masses, the lepton-flavor violating effects from \((Y_\nu^\dagger Y_\nu)_{ij}\) are suppressed by \(\sqrt{\Delta m_{ij}^2}/m_1\), relative to the case of hierarchical neutrino masses, as can be seen in (\ref{llcorrectiondeg}) and Fig.~4. 

\section{Conclusions}
We have systematically studied the sensitivity of $Br(l_{i}\rightarrow l_{j}\gamma)$ and \(\sigma(e^+e^-\rightarrow \tilde{l}_\alpha^+\tilde{l}^-_\beta\rightarrow l^-_i l^+_j + /\!\!\!\!E)\) on the Majorana mass scale $M_{R}$ in new post-LEP mSUGRA scenarios.
Furthermore, we have investigated the correlation of LFV in these low and high energy processes.
In the case of normal neutrino mass hierarchies a measurement of $Br(\mu\rightarrow e\gamma)\approx 10^{-14}$ would imply a value of $M_{R}$ between $3\cdot 10^{11}$~GeV and $4\cdot 10^{14}$~GeV, depending on the mSUGRA scenario. On the other hand, a measurement of $Br(\tau\rightarrow \mu\gamma)\approx 10^{-9}$ would actually determine \(M_R\) with an accuracy of a factor of 2 for a given scenario. 
Moreover, we find that the present bound on \(Br(\mu\rightarrow e \gamma)\) still allows for cross-sections \(\sigma(e^+e^-\rightarrow \mu^-\tau^++2\tilde{\chi}_1^0)\) and \(\sigma(e^+e^-\rightarrow e^-\mu^++2\tilde{\chi}_1^0)\) of the order of 1~fb at c.m. energies of 500 to 800~GeV.

\section*{Acknowledgements}   
This work was supported by the Bundesministerium f\"ur Bildung und Forschung (BMBF, Bonn, Germany) under the contract number 05HT1WWA2.

\end{document}